\documentclass[fleqn,twoside]{article}
\usepackage{espcrc2}

\usepackage{graphicx,psfrag}
\usepackage[figuresright]{rotating}

\newcommand{\AmS}{{\protect\the\textfont2
  A\kern-.1667em\lower.5ex\hbox{M}\kern-.125emS}}

\title{Radio detection of cosmic ray air showers with LOPES}

\author{
T.~Huege\address[addr1]{IK, Forschungszentrum Karlsruhe,76021 Karlsruhe, Germany},
W.D.~Apel\addressmark[addr1], 
T.~Asch\address[addr2]{IPE, Forschungszentrum Karlsruhe, 76021 Karlsruhe, Germany},
A.F.~Badea\addressmark[addr1],
L.~B\"ahren\address[addr3]{ASTRON, 7990 AA Dwingeloo, The Netherlands},
K.~Bekk\addressmark[addr1],
A.~Bercuci\address[addr4]{Nat. Inst. of Physics and Nuclear Eng., 7690 Bucharest, Romania},
M.~Bertaina\address[addr5]{Dipartimento di Fisica Generale dell' Universita, 10125 Torino, Italy},
P.L.~Biermann\address[addr6]{Max-Planck-Institut f\"ur Radioastronomie, 53121 Bonn, Germany},
J.~Bl\"umer\addressmark[addr1]\address[addr7]{IEKP, Universit\"at Karlsruhe, 76021 Karlsruhe, Germany},
H.~Bozdog\addressmark[addr1],
I.M.~Brancus\addressmark[addr4],
S.~Buitink\address[addr8]{Dpt. Astrophysics, Radboud Univ., 6525 ED Nijmegen, The Netherlands},
M.~Br\"uggemann\address[addr9]{Fachbereich Physik, Universit\"at Siegen, 57072 Siegen, Germany},
P.~Buchholz\addressmark[addr9],
H.~Butcher\addressmark[addr3],
A.~Chiavassa\addressmark[addr5],
F.~Cossavella\addressmark[addr7],
K.~Daumiller\addressmark[addr1],
F.~Di~Pierro\addressmark[addr5],
P.~Doll\addressmark[addr1],
R.~Engel\addressmark[addr1],
H.~Falcke\addressmark[addr3]\addressmark[addr6]\addressmark[addr8],
H.~Gemmeke\addressmark[addr2],
P.L.~Ghia\address[addr10]{Istituto di Fisica dello Spazio Interplanetario, INAF, 10133 Torino, Italy}, 
R.~Glasstetter\address[addr11]{Fachbereich C $-$ Physik, Uni Wuppertal, 42097 Wuppertal, Germany},
C.~Grupen\addressmark[addr9],
A.~Hakenjos\addressmark[addr7],
A.~Haungs\addressmark[addr1],
D.~Heck\addressmark[addr1],
J.R.~H\"orandel\addressmark[addr7],
A.~Horneffer\addressmark[addr8],
P.G.~Isar\addressmark[addr7],
K.H.~Kampert\addressmark[addr11],
Y.~Kolotaev\addressmark[addr9],
O.~Kr\"omer\addressmark[addr2],
J.~Kuijpers\addressmark[addr8],
S.~Lafebre\addressmark[addr8],
H.J.~Mathes\addressmark[addr1],
H.J.~Mayer\addressmark[addr1],
C.~Meurer\addressmark[addr1],
J.~Milke\addressmark[addr1],
B.~Mitrica\addressmark[addr4],
C.~Morello\addressmark[addr10],
G.~Navarra\addressmark[addr5],
S.~Nehls\addressmark[addr1],
A.~Nigl\addressmark[addr8],
R.~Obenland\addressmark[addr1],
J.~Oehlschl\"ager\addressmark[addr1],
S.~Ostapchenko\addressmark[addr1],
S.~Over\addressmark[addr9],
M.~Petcu\addressmark[addr4],
J.~Petrovic\addressmark[addr8],
T.~Pierog\addressmark[addr1],
S.~Plewnia\addressmark[addr1],
H.~Rebel\addressmark[addr1],
A.~Risse\address[addr12]{Soltan Institute for Nuclear Studies, 90950 Lodz, Poland},
M.~Roth\addressmark[addr1],
H.~Schieler\addressmark[addr1],
O.~Sima\addressmark[addr4],
K.~Singh\addressmark[addr8],
M.~St\"umpert\addressmark[addr7],
G.~Toma\addressmark[addr4],
G.C.~Trinchero\addressmark[addr10],
H.~Ulrich\addressmark[addr1],
J.~van~Buren\addressmark[addr1],
W.~Walkowiak\addressmark[addr9],
A.~Weindl\addressmark[addr1],
J.~Wochele\addressmark[addr1],
J.~Zabierowski\addressmark[addr12],
J.A.~Zensus\addressmark[addr6],
D.~Zimmermann\addressmark[addr9]\\
\vspace*{0.4cm}The LOPES Collaboration\vspace*{0.2cm}}

\begin{document}

\begin{abstract}
In the last few years, radio detection of cosmic ray air showers has experienced a true renaissance, becoming manifest in a number of new experiments and simulation efforts. In particular, the LOPES project has successfully implemented modern interferometric methods to measure the radio emission from extensive air showers. LOPES has confirmed that the emission is coherent and of geomagnetic origin, as expected by the geosynchrotron mechanism, and has demonstrated that a large scale application of the radio technique has great potential to complement current measurements of ultra-high energy cosmic rays. We describe the current status, most recent results and open questions regarding radio detection of cosmic rays and give an overview of ongoing research and development for an application of the radio technique in the framework of the Pierre Auger Observatory.
\vspace{1pc}
\end{abstract}

\maketitle

\section{Introduction}

About 40 years ago, Jelley et al. \cite{Jelley1965} measured pulsed radio emission originating from extensive air showers (EAS) for the first time. As the radio technique proved to be too difficult to handle with the technical limitations of the 1960s and 1970s, however, the interest in radio detection of cosmic rays diminished completely within the following decade. It was only recently, now having powerful digital technology at our disposal, that the concept of radio detection of cosmic rays experienced its renaissance \cite{FalckeGorham2003}. By now a number of new projects dedicated to the measurement of radio emission from EAS, most prominently the LOPES project \cite{HornefferArena2005,Horneffer2006} and the CODALEMA project \cite{Codalema}, have been established.

The radio technique for measuring cosmic rays has a number of merits in its own right: it can watch for EAS with nearly 100\% duty cycle, even in populated areas, and it measures a bolometric signal that is only very slightly attenuated in the atmosphere, thus allowing the observation of highly inclined showers. But naturally, the most interesting application is to combine the technique with other detection methods, in particular ground-based particle detector arrays and air fluorescence measurements. Each of these techniques yields different observables, and a combination of the methods allows so-called ``hybrid detection'' of cosmic rays, yielding much more information than the individual techniques alone.

In this article, we review the goals, the status and the results so far gathered within the LOPES project followed by an outlook on the application of the radio technique on large scales for the measurement of ultra-high energy cosmic rays.

\section{The LOPES project}

The LOPES project was initiated in 2001, employing a {\bf LO}FAR\footnote{LOFAR is a revolutionary digital radio interferometer for the 10-200 MHz range being set up in the Netherlands.} {\bf P}rototyp{\bf e S}tation to detect radio emission from EAS in coincidence with the well-established KASCADE \cite{KASCADE2003} particle detector experiment. The goals of the LOPES project are (i) to deliver the ``proof of principle'' for radio detection of cosmic rays with modern interferometric methods, (ii) to study and calibrate the radio emission in the energy regime up to $\sim10^{18}$~eV, and (iii) to develop and optimise the radio technique for large scale application at ultra-high energies. To achieve these goals the LOPES project combines both experimental activities and dedicated theoretical efforts.

\vspace{-0.2cm}\subsection{Simulations}\vspace{0.25cm}

Historical studies of the theory of radio emission from EAS already pointed to a geomagnetic process as the dominant emission mechanism \cite{Allan1971}. These studies, however, were not detailed enough to serve as a basis for the interpretation of LOPES experimental data. We therefore started our own simulations of radio emission from EAS, concentrating on the dominating geomagnetic emission. The scheme in which this mechanism is described is that of ``coherent geosynchrotron radiation'' \cite{FalckeGorham2003} from shower electrons and positrons that are deflected in the earth's magnetic field.

In a first step, we studied the emission with a frequency-domain analytical model, analysing in particular the important coherence effects arising because the length scales present in the air shower are of the same order as the observing wavelength in the frequency range of LOPES \cite{HuegeFalcke2003}. In a second step, we developed a detailed time-domain Monte Carlo simulation of the emission process \cite{HuegeFalcke2005a} predicting the radio emission and its dependence on different air shower parameters with unprecedented detail \cite{HuegeFalcke2005b}.

\begin{figure}[tb]
   \psfrag{Eomegaew0muVpmpMHz}[c][t]{$\left|E_{\omega}(2\pi\nu_{0})\right|$~[$\mu$V~m$^{-1}$~MHz$^{-1}$]}
   \centering
   \includegraphics[width=5.5cm,angle=270]{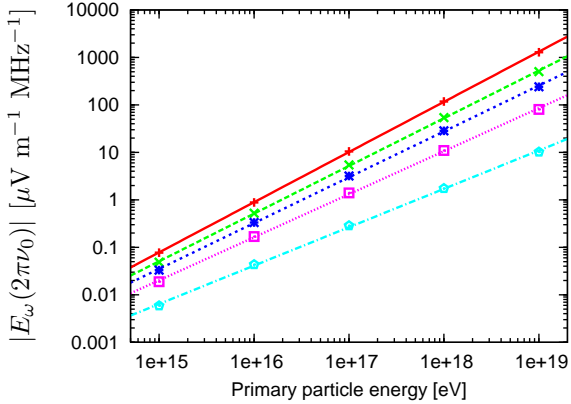}
   \vspace{-0.9cm}
   \caption[Scaling with primary particle energy]{
   \label{fig:energydependence}
   Scaling of the field strength emitted at $\nu_{0}=10$~MHz by a vertical EAS as a function of primary particle energy \cite{HuegeFalcke2005b}. From top to bottom: 20~m, 100~m, 180~m, 300~m and 500~m to the north from the shower centre. The field strength scales as a power-law with index close to unity, as expected for coherent emission.\vspace{-0.6cm}
   }
   \end{figure}

\begin{figure}[tb]
   \psfrag{Eomegaew0muVpmpMHz}[c][t]{$|E_{\omega}(2\pi\nu)|$~[$\mu$V~m$^{-1}$~MHz$^{-1}$]}   
   \psfrag{nu0MHz}[c][b]{$\nu$~[MHz]}
   \centering
   \includegraphics[width=5.5cm,angle=270]{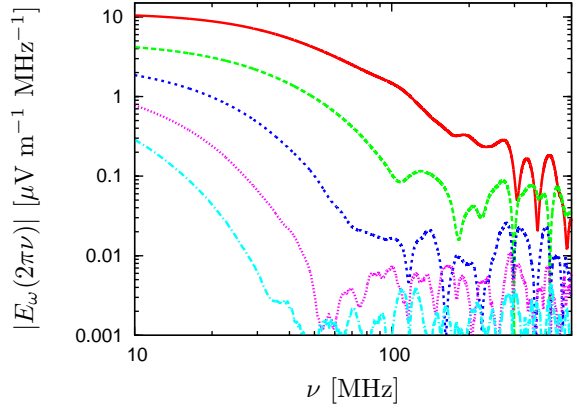}
   \vspace{-0.9cm}
   \caption[Spectra of a vertical air shower]{
   \label{fig:spectravertical}
   Frequency spectra of a 10$^{17}$~eV vertical air shower at various distances to the north from the shower centre \cite{HuegeFalcke2005b}: From top to bottom: 20~m, 140~m, 260~m, 380~m and 500~m. ``Noise'' in the incoherent high-frequency regime stems from the simplified air shower model and statistics.\vspace{-0.6cm}
   }
   \end{figure}

\begin{figure}[tb]
   \centering
   \includegraphics[width=7.4cm]{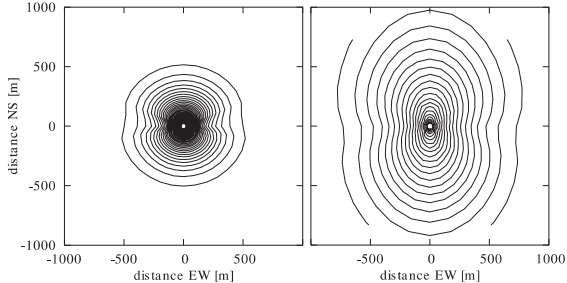}
   \vspace{-1.2cm}
   \caption[Radio footprints]{
   \label{fig:footprint}
   10~MHz radio footprints of a vertical $10^{17}$~eV air shower (left) and a 45$^{\circ}$ inclined $10^{17}$~eV air shower coming from the south (right) \cite{HuegeFalcke2005b}. The signal increases by 0.25 $\mu$V~m$^{-1}$~MHz$^{-1}$ per contour line. Inclined showers have a much larger radio footprint, making them particularly suitable for radio detection.\vspace{-0.6cm}
   }
   \end{figure}

Important predictions of these simulations are that (i) the radio signal should scale approximately linearly with the primary particle energy in the coherent frequency range (see Fig.\ \ref{fig:energydependence}), (ii) the frequency spectrum declines rather steeply to high frequencies, making low frequencies favourable for detection (see Fig.\ \ref{fig:spectravertical}), (iii) the electric field strength should decrease exponentially with radial distance to the shower centre, (iv) the dependence on the strength and geometry of the magnetic field is much more subtle than intuitively expected and mostly shows up in the polarisation characteristics of the radio signal, and (v) inclined showers exhibit a much larger radio ``footprint'', making them especially favourable for radio detection (see Fig.\ \ref{fig:footprint}). The essence of these elaborate simulations is summarised by a parametrisation formula that can be a useful tool to quickly estimate the strength of radio emission for a particular situation \cite{HuegeFalcke2005b}.

Recently, we have developed a next-generation Monte Carlo code, replacing the parametrised description of air shower characteristics with a highly realistic CORSIKA \cite{Heck1998} based air shower model. First results are presented in  \cite{HuegeArena2006}.

\subsection{Experiment}\vspace{0.25cm}

LOPES uses so-called ``inverted V'' dipole antennas, each on their own sensitive to a large fraction of the sky, to measure radio signals in the 40--80 MHz band. A readout of the LOPES antennas is triggered whenever the KASCADE array registers a large (approx.\ $>10^{16}$~eV) air shower. This coincident measurement of the particle and radio components of an EAS is imperative for the unambiguous association of radio detections to air showers and, consequently, air shower parameters on an event-to-event basis.

Using the air shower geometry as reconstructed by KASCADE, the radio data gathered by the individual LOPES antennas can then be time-shifted and correlated appropriately in an offline analysis to let the array of antennas work as an interferometer looking into the specific direction of the air shower. This process is called ``beam-forming'' and is essential for increasing the signal-to-noise ratio of the radio measurements. For details of the technical aspects of LOPES and data analysis procedures we refer the reader to \cite{Horneffer2006}.

   \begin{figure}[tb]
   \centering
   \includegraphics[width=7.0cm]{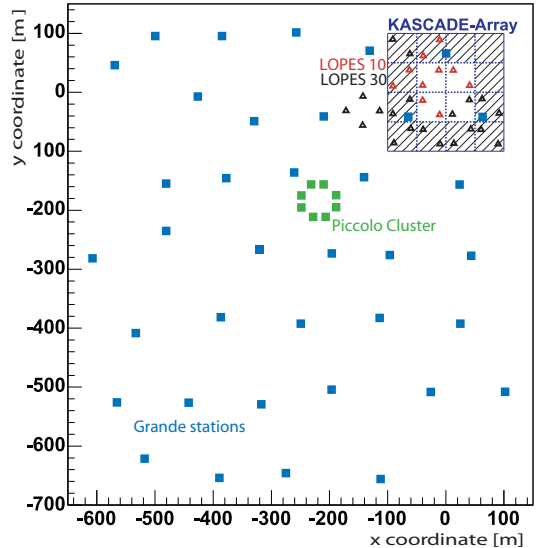}
   \vspace{-0.9cm}
   \caption[LOPES layout]{
   \label{fig:layout}
   Layout of the LOPES10 and LOPES30 array configurations relative to the KASCADE-Grande surface detector components.\vspace{-0.6cm}
   }
   \end{figure}

In the initial LOPES setup, 10 linearly east-west polarised antennas were situated inside the KASCADE array of the KASCADE-Grande experiment \cite{Grande2004} (red triangles in Fig.\ \ref{fig:layout}). This configuration, hereafter called LOPES10, collected 7 effective months of data in the year 2004. The results derived from these measurements are presented in section \ref{sec:results}.

After the measurements with 10 antennas had gained good statistics, the array was rearranged and extended to 30 linearly east-west polarised antennas situated mostly inside the KASCADE array (black triangles in Fig.\ \ref{fig:layout}). This configuration, hereafter called LOPES30, offers larger baselines (e.g., for a better angular resolution), a larger collecting area, and allows analyses with independent ``sub-arrays'' of antennas, e.g., to measure the lateral profile of the radio emission on a per-event basis. The most important enhancement from LOPES10 to LOPES30 was, however, that the LOPES30 array has been fully calibrated with an external reference source, allowing to finally address the 40-year-old question of the absolute field strength of radio emission from EAS, in comparison with the simulation predictions. In addition, a detailed environmental monitoring allows to correlate the radio data to many external variables, including the static atmospheric electric field at ground level \cite{IsarArena2006}.

Currently, LOPES is being reconfigured to make dual-polarisation measurements. These will allow a much better analysis of the angular correlations in the radio data, and a clear verification of the geomagnetic emission process. An additional triggering by the Grande stations of the KASCADE-Grande experiment is also planned.

\subsection{LOPES10 Results}\vspace{0.25cm}\label{sec:results}

The first analysis of LOPES10 data concentrated on a limited number of very energetic events, selected by cuts on high (truncated) muon number and high electron number as provided by KASCADE. The results of this analysis \cite{Nature2005} were that (i) there is an unambiguous association of pulsed radio emission with air showers, and the KASCADE-reconstructed direction is consistent with the source of the radio emission, (ii) the radio field strength correlates very well with the muon number, which in turn correlates well with the primary particle energy, and (iii) the radio field strength correlates clearly with the angle between shower axis and earth's magnetic field (hereafter called geomagnetic angle). With this analysis, the ``proof of principle'' for radio detection of EAS with modern interferometric methods had been achieved. Additionally, the observed angular correlation with the geomagnetic field proved that a large fraction of the emission must be produced by a geomagnetic effect.

\begin{figure}[tb]
   \centering
   \includegraphics[height=7.0cm,angle=270]{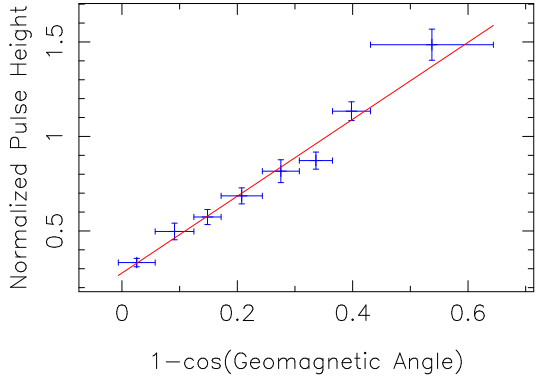}
   \vspace{-0.8cm}
   \caption[Geomagnetic angle correlation]{
   \label{fig:horneffgeomag}
   Correlation of the radio pulse field strength (normalised by the assumed correlation on truncated muon number) with the angle between shower axis and geomagnetic field \cite{Horneffer2006}.\vspace{-0.6cm}
   }
   \end{figure}

\begin{figure}[tb]
   \centering
   \includegraphics[height=7.0cm,angle=270]{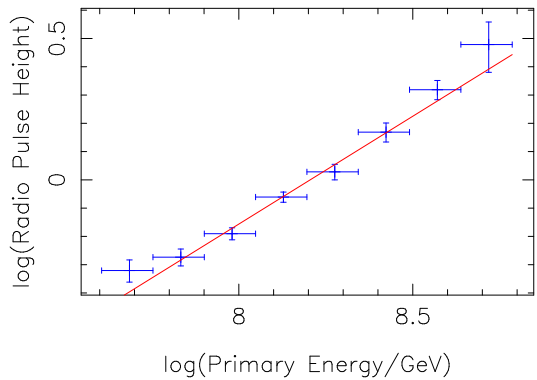}
   \vspace{-0.8cm}
   \caption[Energy correlation]{
   \label{fig:horneffenergy}
   Correlation of the radio pulse field strength (normalised by the assumed correlation on geomagnetic angle) with the primary particle energy as reconstructed by KASCADE \cite{Horneffer2006}.\vspace{-0.6cm}
   }
   \end{figure}

A repetition of this analysis with much higher statistics and for multiple data sets selected with different cuts was repeated in \cite{Horneffer2006}. This elaborate analysis confirmed the earlier findings and constrained the correlations, e.g.\ on geomagnetic angle (see Fig.\ \ref{fig:horneffgeomag}), much more precisely. The results have been summarised in a parametrisation formula describing the measured radio emission's dependence on the major air shower parameters. The scaling with primary particle energy is confirmed to be almost linear (see Fig.\ \ref{fig:horneffenergy}), as expected from theory for coherent emission. Also, the electric field turns out to decrease exponentially with radial distance from the shower axis, again as expected from the simulations. A direct comparison of the angular correlations with the simulations is difficult, however, because LOPES10 only measured the east-west linear polarisation component, thus missing a significant fraction of the emission depending on the air shower geometry. Also, the field strength of LOPES10 data is not absolute calibrated, a problem which will be solved with LOPES30 data.

Both experimental data and simulations show similarities, but also differences, to the parametrisation of the historical experimental data given in \cite{Allan1971}. As the historical measurements were done with a very different setup and the documentation of the data is not in all cases unambiguous, the differences are, however, not surprising.

   \begin{figure}[tb]
   \centering
   \includegraphics[width=7.4cm]{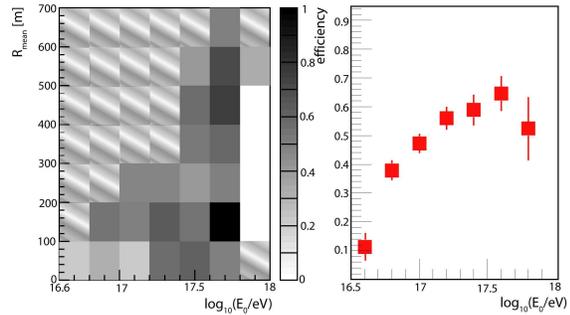}
   \vspace{-1.3cm}
   \caption[Efficiency of radio detection]{
   \label{fig:efficiency}
   Radio detection efficiency as a function of mean radial distance between shower core and antennas as well as primary particle energy (left) and primary particle energy alone (right) \cite{Badea2006}. Detection efficiencies are for purely east-west polarised antennas.\vspace{-0.6cm}
   }
   \end{figure}

Another analysis has been carried out using the Grande stations (instead of the KASCADE array) to reconstruct the air shower parameters \cite{Badea2006}. The air showers selected for this analysis have their core on average at much larger distances to the radio antennas than the ones selected in the previously described analyses. The most important results of this analysis are (i) LOPES does detect radio emission from EAS at distances as large as 500~m from the shower core for energies well below $10^{18}$~eV, an important result for a possible use of the radio technique on large scales, (ii) the radio pulse field strength depends significantly on the accuracy of the input parameters (core position, shower axis), i.e., radio measurements have the potential to constrain these parameters with high precision and thus help in the event reconstruction of classical particle detector arrays, (iii) an exponential decrease of the radio signal with radial distance to the core seems to describe the data also to large distances, and (iv) LOPES reaches a detection efficiency of approx.\ $>50$\% starting at approx.\ $10^{17}$~eV (see Fig.\ \ref{fig:efficiency}), which is encouraging when taking into account that LOPES10 misses up to half of the emission because it measures only the east-west linear polarisation of the radiation.

   \begin{figure}[tb]
   \centering
   \includegraphics[width=7.0cm]{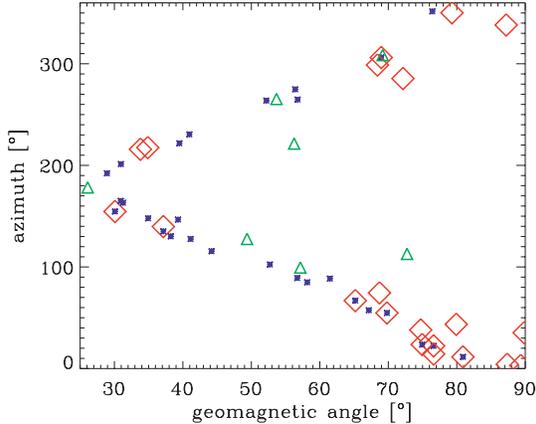}
   \vspace{-0.8cm}
   \caption[Inclined showers angular correlations]{
   \label{fig:jelenaangles}
   Inclined air showers distribution in zenith angle and geomagnetic angle \cite{Petrovic2006}. Blue crosses: events fully reconstructed by KASCADE, green triangles: events only geometrically reconstructed by KASCADE, red rhombs: events detected in radio. No showers from 90$^{\circ}$ and 270$^{\circ}$ azimuth angles (denoting east and west) are detected in radio.\vspace{-0.6cm}
   }
   \end{figure}

As predicted by the simulations, inclined air showers pose a particularly interesting target for radio observations. On the one hand, their radio footprint is expected to be large, and on the other hand, inclined showers provide a larger ``lever arm'' to analyse the radio field strength correlations on zenith, azimuth and geomagnetic angles. Also, nearly horizontal air showers can be induced by neutrino interactions in the atmosphere, making them a particularly interesting field of research. In \cite{Petrovic2006} we have analysed a selection of air showers with zenith angles over 50$^{\circ}$, an angular range where flat particle detectors such as those of KASCADE become inefficient. The analysis has shown that (i) LOPES does detect very inclined air showers up to approx.\ 80$^{\circ}$ zenith angle, (ii) the radio detection efficiency rises with increasing zenith angle, and (iii) angular effects are indeed more pronounced in this zenith angle range. Figure \ref{fig:jelenaangles} shows the KASCADE triggered and radio detected events in the plane of azimuth angle versus geomagnetic angle. While statistics are still somewhat low, one can see two effects: First, there is a pronounced north-south asymmetry (more radio detections for showers coming from the south with large geomagnetic angle than from the north with small geomagnetic angle). Such an asymmetry was also observed by the historical experiments and can be explained by the simulations when threshold effects play a role, as is certainly the case for LOPES. Second, there is a pronounced gap with no radio detections at azimuth angles of approx.\ 90$^{\circ}$ and 270$^{\circ}$, i.e., there are no radio detections of showers coming from east or west. As LOPES10 only measured the east-west linear polarisation of the emission, this is consistent with the simulation prediction that showers coming from east or west should be mostly north-south polarised. Additionally, antenna gain effects could play a role.

   \begin{figure}[tb]
   \centering
   \includegraphics[width=7.0cm]{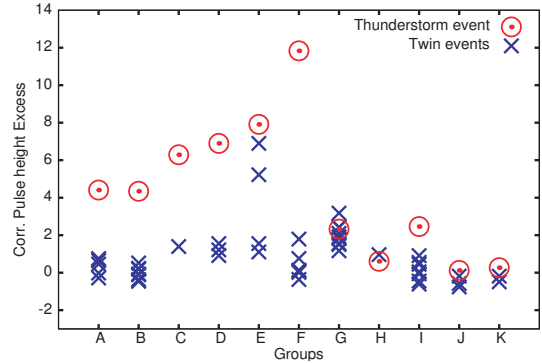}
   \vspace{-0.8cm}
   \caption[Thunderstorm effects]{
   \label{fig:stijntwins}
   Enhancement of the radio emission during thunderstorm conditions in comparison with events with similar air shower parameters recorded during fair weather \cite{Buitink2006}.
   \vspace{-0.6cm}
   }
   \end{figure}

An analysis of radio events measured during thunderstorm conditions \cite{Buitink2006} revealed that radio emission from EAS seems to be amplified by the strong (of order kV/cm) atmospheric electric fields present in thunderstorm clouds. Figure \ref{fig:stijntwins} compares the field strengths of events during thunderstorms with that of events with similar air shower parameters measured during fair weather (``twin events''). In a number of cases, a clear enhancement of the radio emission is visible. It also becomes clear that amplification of radio emission only occurs when very high electric field strengths are present in the atmosphere. The fair weather electric field (of order 10 V/m) does not influence the radio emission from EAS. These effects are currently investigated in more detail using extensive simulations and LOPES30 data.

\section{The current understanding}

As of today, some properties of radio emission from EAS are well understood: The emission is clearly coherent as visible from the approximately linear scaling of field strength with energy seen in both experimental data and simulations \cite{Horneffer2006,Nature2005}. Also, the fact that there is a pronounced correlation between the radio field strength and the geomagnetic angle indicates that a geomagnetic emission mechanism must be responsible for the dominant fraction of the emission \cite{Horneffer2006,Nature2005,Petrovic2006}.

Some theoretical predictions seem to be confirmed but need to be investigated further with additional data: The radial dependence of the radio field strength is described well by an exponential decrease both in experimental data and simulations \cite{Allan1971,HuegeFalcke2005b,Badea2006,CodalemaExponential}. Also, as expected from theory, inclined air showers can be detected very well with the radio technique \cite{Petrovic2006}.

Some important questions have yet to be answered: Are the polarisation characteristics of the emission and the detailed angular correlations as expected by theory \cite{HuegeFalcke2005b}? Results from the analysis of inclined air showers \cite{Petrovic2006} are consistent with the predicted signal polarisation, but only dedicated dual-polarisation measurements as are currently being prepared with LOPES will answer these questions definitely. What is the absolute strength of the emission, and is it consistent with the values predicted by theory \cite{HuegeFalcke2005b}? The LOPES30 calibrated measurements will answer this long-standing question soon. What is the scale parameter of the radio emission's exponential radial decay, and how does it relate to the air shower geometry? LOPES30 will allow a much more detailed investigation of the per-event lateral radio profile than LOPES10. And finally, is the radio emission sensitive to the primary mass of the EAS \cite{HuegeICRC2005} and how can this be exploited? This question and many others can be investigated with the new CORSIKA-based Monte Carlo code \cite{HuegeArena2006}.

\section{The way to high energies}

The radio technique has high potential for application on large scales, and, consequently, at high energies. In particular, its ability to measure with nearly 100\% duty cycle is a strong asset for combining it with surface detector arrays to take advantage of hybrid detection of EAS. One application of the radio technique on a much larger scale than LOPES or CODALEMA will be the use of LOFAR to study cosmic rays.

Another vision is to use the radio technique for hybrid detection of ultra-high energy cosmic rays in the framework of the Pierre Auger Observatory (PAO). It is clear, however, that an application on this scale has requirements very different from those of already existing experiments. For once, it will be necessary to distribute the antennas on a grid much larger than that of the current experiments. A grid spacing of order 500~m or more is probably needed to make a large-scale radio array cost-effective. As a direct consequence, the organisation of such a radio array will have to be decentralized with mostly autonomous detectors (low power-consumption, ...) and wireless data transfer. The possibility to self-trigger on the radio signal only would also be highly desirable (although not necessary in case of the PAO), and the feasibility of this is related to the grid spacing and the environmental noise levels (which luckily are much lower in Argentina than in environments such as the Forschungszentrum Karlsruhe).

A number of groups with expertise in radio detection of EAS are currently developing the necessary techniques, evaluating alternative antenna designs, and developing electronics and data links. Some small test cells are going to be set up inside the Pierre Auger Observatory starting September 2006 in order to evaluate their performance in the Argentinian noise environments.

The experience and results gained --- and to be gained --- from LOPES at energies up to approx.\ $10^{18}$~eV are invaluable in their own right for the adaptation of the radio technique to large scales. In addition, LOPES has developed a new kind of self-triggering array called LOPES$^{\mathrm{STAR}}$. It was conceived to meet the requirements of a large scale decentralized array and is using a different type of electronics and antennas. Detectors of the LOPES$^{\mathrm{STAR}}$ type are going to be installed as part of the test cells in the Pierre Auger Observatory. At the same time, they will be operated and further developed in the environment of the original LOPES array, which continues to provide a unique test bed for the study of radio emission from EAS.

\section{Conclusions}

In the last few years, radio detection of cosmic ray air showers has once again become a very active field of research. LOPES, as one of the projects studying radio emission from EAS with modern digital technology, has made important contributions. Radio emission from EAS is proven to be coherent at 40--80~MHz, and the field strength correlation with geomagnetic angle strongly indicates that the radiation is of dominantly geomagnetic origin. Aspects such as the absolute field strength, the detailed angular correlations and the scale radius of the lateral dependence will soon be analysed with the new LOPES30 data. Our next-generation Monte Carlo code calculating the radio emission based on highly realistic CORSIKA-based air showers will also allow many new studies with unprecedented detail. Unlike 40 years ago, the study of radio emission today is making good progress.

A possible application of the radio technique on large scales for hybrid detection of ultra-high energy cosmic rays is now being studied in the framework of the Pierre Auger Observatory. These activities are still in a relatively early phase, but with the experience gained in the LOPES and CODALEMA experiments, the challenges to unlock the great potential of the radio technique on large scales can be tackled. 

%


\vspace{0.4cm}

{\small\noindent Acknowledgements: LOPES was supported by the German Federal Ministry of Education and Research (Verbundforschung Astroteilchenphysik) and is part of the research programme of the Stichting voor Fundamenteel Onderzoek der Materie (FOM), which is financially supported by the Nederlandse Organisatie voor Wetenschappelijk Onderzoek (NWO). The KASCADE-Grande experiment is supported by the German Federal Ministry of Education and Research, the MIUR of Italy, the Polish Ministry of Science and Higher Education and the Romanian National Academy for Science, Research and Technology.}

\end{document}